\documentclass[prl,twocolumn,showpacs,preprintnumbers]{revtex4}
\usepackage{amsfonts}
\usepackage{amsmath}
\usepackage{amssymb}
\usepackage{graphicx}
\usepackage{dcolumn}
\usepackage{bm}
\usepackage{mathcomp}
\usepackage{subfigure}
\usepackage{lipsum}

\begin{document}

\title{Fast closed-loop optimal control of ultracold atoms in an optical lattice}

\author{S.~Rosi$^{1}$} \author{A.~Bernard$^{1}$} \author{N.~Fabbri$^{1}$} \author{L.~Fallani$^{1,2}$} \author{C.~Fort$^{1}$} \author{M.~Inguscio$^{1,2}$}
\affiliation{
$^1$LENS and Dipartimento di Fisica e Astronomia, Universit\`{a} di Firenze and INO-CNR, 50019 Sesto Fiorentino, Italy\\
$^2$QSTAR Center for Quantum Science and Technology, Largo Enrico Fermi 2, I-50125 Arcetri, Italy
}
\author{T.~Calarco$^3$}
\author{S.~Montangero$^3$}
\affiliation{
$^3$Institut f\"ur Quanteninformationsverarbeitung, Universit\"at Ulm, D-89069 Ulm, Germany
}

\date{\today}
\begin{abstract}
We present experimental evidence of the successful closed-loop optimization of the dynamics of cold atoms in an optical lattice. We optimize the loading of an ultracold atomic gas minimizing the excitations in an array of one-dimensional tubes (3D-1D crossover) and we perform an optimal crossing of the quantum phase-transition from a Superfluid to a Mott-Insulator in a three-dimensional lattice. In both cases we enhance the experiment performances with respect to those obtained via adiabatic dynamics, 
effectively speeding up the process by more than a factor three while improving the quality of the desired transformation. 
\end{abstract}

\pacs{67.85.Hj, 03.67.-a, 05.30.Rt}

\maketitle

In the last decade, the implementation of quantum simulators with cold atoms has experienced remarkable expansion~\cite{BlochNP12}. The latest developments in the field have made now possible to experimentally investigate Fermi and Bose ultracold gases in many different setups~\cite{ZwergerRMF2008}. 
Optical potentials have given access to the simulation of the ground-state physics and the dynamics of some of the most important lattice models: Hubbard and spin models have been successfully implemented~\cite{EsslingerArxiv1212,BlochNAT02,EsslingerNAT2008,RoschSCI08,simonNat11}. Including artificial disorder enables the study of ubiquitous phenomena like Anderson localization~\cite{roatiN08}. Recently, improved experimental techniques allowed for the acquisition of unprecedented single-atom resolved images
and the coherent control of single spins~\cite{shersonN10,weitenbergN11} paving the way for the next generation of experiments. 
Novel and more challenging ideas have been proposed to exploit the potential of quantum simulators to study artificial gauge fields related 
to quantum Hall physics~\cite{gerbierNJP10}, the physics of complex quantum systems~\cite{montangeroEPL09} and gauge theories~\cite{banerjeePRL12}.
The path towards new experiments of increasing complexity is conditional on the development of better and more precise experimental techniques, 
to achieve increased control on the system under investigation. The necessary steps to be taken are mostly related to technological and experimental development, 
however recently an important theoretical contribution has been put forward. Indeed, it has been shown that it is possible to exploit 
quantum optimal control to synthesize optimal strategies for correlated quantum many-body dynamics~\cite{doria_PRL11,Caneva_PRA11}, as already known
for few-body or uncorrelated quantum systems~\cite{brif_NJP10, demler, machnes_PRL10,buckerNP11}. Combining numerical simulations and novel 
approaches has enabled optimal control of correlated quantum many-body dynamics and optimal driving of
phase transitions~\cite{doria_PRL11,caneva_pra11_1} and the engineering of many-body entangled and squeezed states~\cite{caneva_NJP12}.
\begin{figure}[t]
\centering
  \includegraphics[width=1\columnwidth]{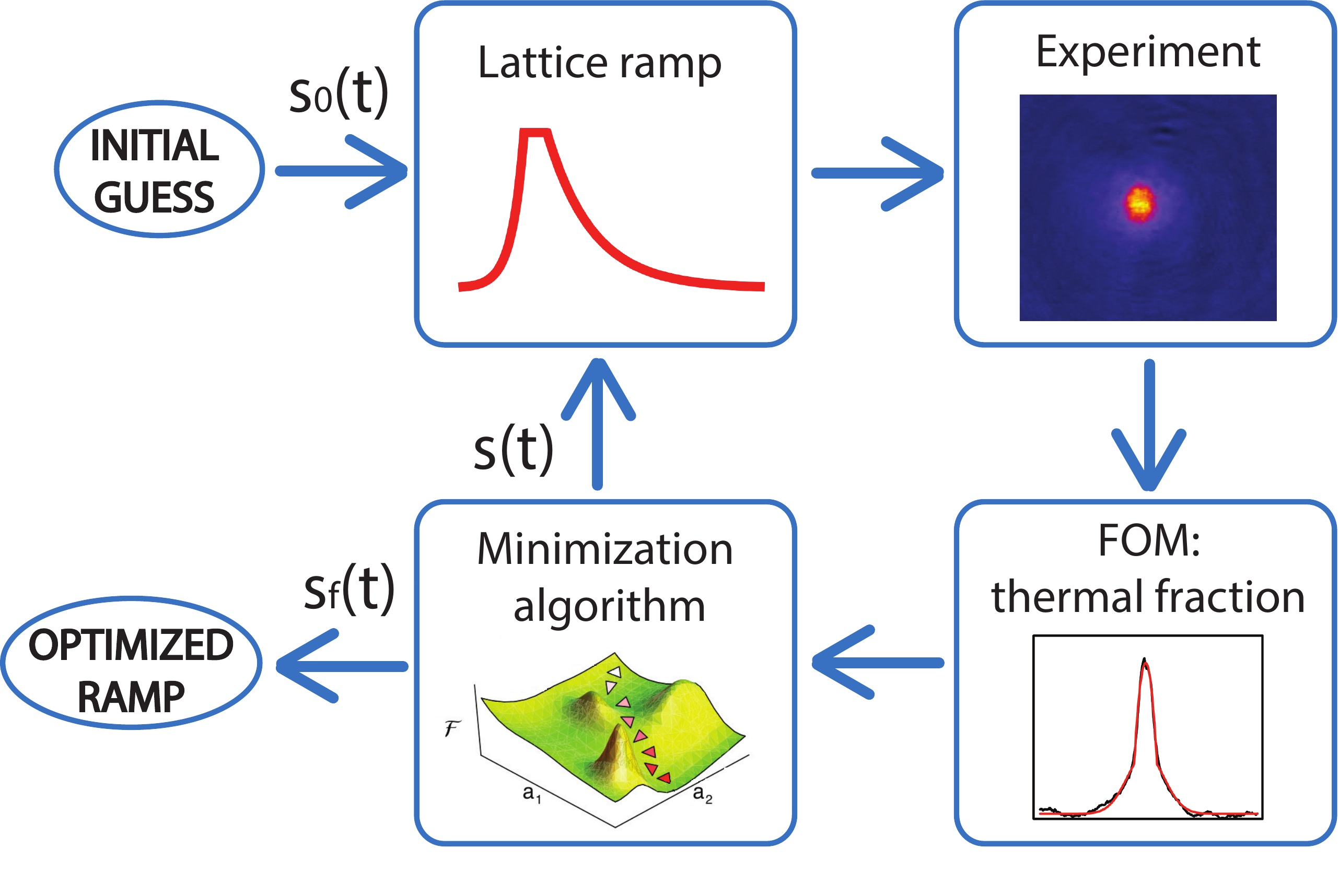}
  \caption{Scheme of closed-loop optimization experiment. The control-field $s\left(t\right)$ describes the temporal dependence of the lattice depth during the loading of the atomic gas in the lattice. An initial guess $s_0(t)$ is chosen: for each experimental run and a Time-Of-Flight image is recorded. From a double-structure fit of the density distribution we extract the Figure Of Merit $F$. The optimization algorithm provides an updated function $s(t)$ and the experiment is repeated until reaching an optimal field $s_f(t)$.}
  \label{fig:Scheme}
\end{figure}
Despite these promising theoretical results, their experimental implementation might be limited by different issues mainly arising from discrepancies between 
theoretical models and experimental realization. Even though optimal control fields are generically robust against noise and imperfections~\cite{montangero_PRL07}, 
it would be desirable to have an optimal control field obtained by means of the most accurate and comprehensive description possible of the system dynamics 
under consideration.  Moreover, there are also cases where open-loop quantum optimal control cannot be applied to a given dynamics as no efficient classical description is available, 
for example in highly-entangled quantum many-body systems in dimension greater than one~\cite{vidalPRL04}. These limitations might  
be overcome by means of closed-loop optimal control, i.e., the application of  optimal control in a loop that includes experiments, where the control fields are updated at each step after 
a direct measurements of a Figure Of Merit (FOM) (see Fig.~\ref{fig:Scheme} for the case discussed in this paper). This approach has the great advantage of taking all sources of uncertainties automatically into account, such as for example limited knowledge of system parameters, errors, and constraints.
Indeed, closed-loop optimization has been successfully applied in different contexts, from NMR to femtosecond laser driving of molecular dynamics~\cite{brif_NJP10}.
Our approach, based on Chopped RAndom Basis (CRAB) optimization, has the additional nice feature of avoiding complex broadband control fields, and 
it is simple to be implemented in the lab. On top of that, CRAB optimization might be used to find simple but unknown optimal control fields that might be exploited in similar or following repetitions of the experiment. Indeed, this has been shown to be the case for the open-loop control of Rydberg atom dynamics where a CRAB optimization on different disordered samples guided the design of a robust optimal control field~\cite{mueller_arixve12}.

In this Letter we apply optimal control theory to give the first experimental demostration of a closed-loop optimal loading of an ultracold atomic sample in optical lattices. We demonstrate the possibility of guiding the system from an initial to a final state through a non-adiabatic transformation, 
such that the final state is reached in a shorter time and with a better fidelity with respect to a slower quasi-adiabatic procedure. We optimize two different fundamental processes that appear in many different cold atoms experiments. We first optimally load a Bose-Einstein condensate (BEC) of $^{87}\text{\text{Rb}}$ atoms in a two-dimensional optical lattice in order 
to produce an array of one-dimensional gases. Afterwards, we optimally 
drive an atomic gas across the quantum phase transition from a superfluid to a Mott insulating phase loading the BEC into a three-dimensional optical lattice. 

{\it Experimental setup -} A degenerate sample of $^{87}\text{Rb}$ is obtained in a hybrid magneto-optical trap realized with the superposition of a quadrupolar magnetic field and a focused red-detuned laser beam~\cite{linPRA2009}. After evaporative cooling, we obtain a BEC of about $2\times 10^{5}$ atoms experiencing an external harmonic potential with cylindrical symmetry ~\cite{noteFreq}. In the slow adiabatic loading procedure, the lattice potential depth $s\left(t\right)$ (expressed in recoil energy units $E_{r}=h^{2}/2m\lambda^2$, where the optical lattice wavelength is $\lambda=830\,\text{nm}$, $h$ is the Planck constant and $m$ is the atomic mass) is increased exponentially from zero to a maximum value $s_{max}$ in a time $\Delta t$ and a time constant $\tau$ according to:
\begin{equation}
\label{eq:ExpRamps}
  s\left(t\right)=s_{max}\frac{1-e^{t/\tau}}{1-e^{\Delta t/\tau}}.
\end{equation}\\
The lattice intensity is controlled by acousto-optic modulators and stabilized at the desired value by a feedback system during the whole transformation. In the optimization tests, we compare the non-adiabatic transformations to an exponential quasi-adiabatic ramp of total duration of $\Delta t_{\text{ad}}=140\,\text{ms}$ and time constant $\tau_{\text{ad}}=30\,\text{ms}$. The latter is representative of typical ramps used in experiments~\cite{gericke2007}. The extimation of the excitations produced in the gas by the non-adiabatic ramp in obtained in the following way. Once the optical lattice has reached the maximum intensity $s_{max}$, it is kept at this value for $5\,\text{ms}$ and then turned off with the time-reversed exponential ramp.
Then, after a thermalization time $\Delta t_{th}$, all the other confining potentials are switched off and the atomic cloud evolves in free space. The energy excess of the lattice gas is quantified by measuring the thermal fraction from absorption imaging after a time of flight (TOF) of $28\,\text{ms}$, when the thermal and condensed parts are well distinguishable~\cite{noteThFr}. The thermal fraction is taken as the FOM to be minimized by the optimization process. Note that the ramping down of the lattice is always adiabatic (exponential decreasing ramp with $\tau_{\text{ad}}$ and $\Delta t_{\text{ad}}$), so that the heating measured at the end of this ramp reflects the heating of the gas in the loading procedure. 

\textit{Optimization} -- A typical quantum optimal control problem is stated as follows:
given a system described by a Hamiltonian function of a time-dependent control field $s(t)$, i.e. $H = H(s(t))$,  we aim to extremize a given FOM $F(s(t))$ after the time evolution driven by $s(t)$ in the time interval $t \in [0,\Delta t]$. CRAB optimization is a strategy that solves the aforementioned problem starting with an initial guess $s_{0}\left(t\right)$ and looks for the optimal control field that extremizes the FOM over all the functions of the form $s(t) = s_0(t) g(t)$ where the correction $g(t)$ is defined as a truncated expansion in some given 
basis functions. In particular here we assume
\begin{equation}
\label{CRAB}
g\left(t\right)= 
\frac{1+ \sum_{j=1}^{n_f} \left(a_j \sin(2 \pi \nu_j t) + b _j \cos (2 \pi \nu_j t)\right)} {1+  \sum_{j=1}^{n_f} 
 \left(a_j \sin(2 \pi \nu_j \Delta t) +
 b _j \cos (2 \pi \nu_j \Delta t)\right)},
\end{equation}
for some (small) number $n_f$ of components with frequencies $ \nu_j$.  The optimization problem is then recast as an extremization  of a multi-variable function 
$F\left(\{h_{j}\}\right)$ of the variables $h_j =(a_j, b_j)$. Here the minimization is performed by means of a Simplex minimization algorithm~\cite{doria_PRL11}.

In optical lattice experiments, this problem typically appears as one aims to prepare a given state varying the optical lattice intensity $s\left(t\right)$. Indeed, our goal is to prepare the system ground-state thus minimizing the residual excitations in the final state. In particular, first we optimize the loading of a BEC in a two-dimensional optical lattice as a test of our closed-loop experiment. Then, we exploit the CRAB method to optimize the Superfluid-Mott insulator quantum phase-transition. 
The experimental procedure of closed-loop optimization is shown in Fig.\ref{fig:Scheme}: we define an initial guess $s_0(t)$ and implement the experimental sequence described above, at the end of which a TOF image is recorded. An automated fitting procedure results in a measurement 
of the final thermal fraction of the sample $\text{TF}=\text{N}_{th}/\text{N}_{tot}$ to be compared with the initial thermal fraction $\text{TF}_{i}$. Their ratio defines the FOM $F=\text{TF}/\text{TF}_{i}$ we minimize. With this information the minimization subroutine implements a search in the parameters space that defines an 
updated loading ramp $s(t)$. The loop is then closed and the process repeated until convergence or when the given desired precision has been reached. 
\begin{figure}[th!]
\centering
\subfigure
{
  \includegraphics[width=1.0\columnwidth]{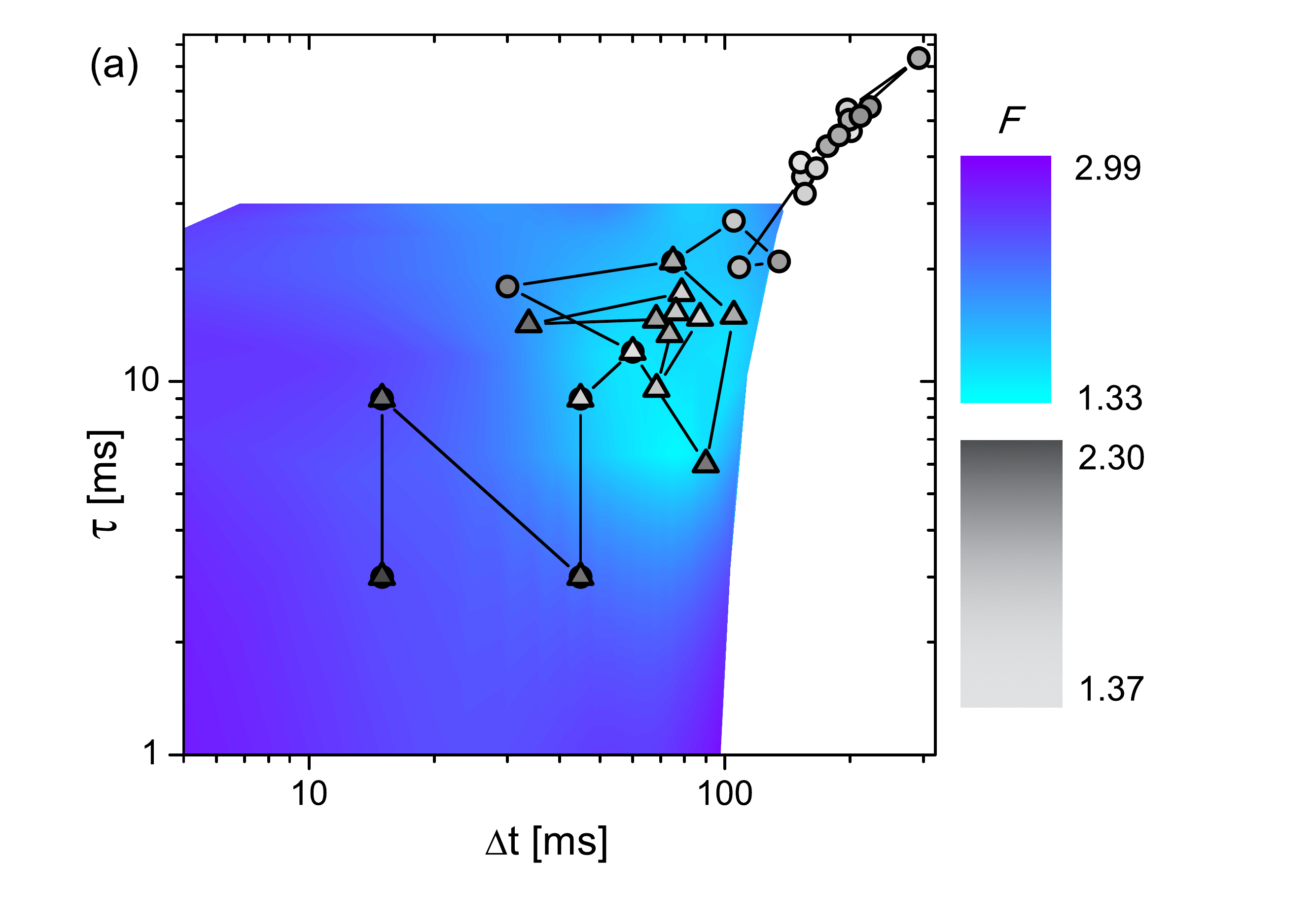}  
  \label{fig:2DolRUN12params}
}
\subfigure
{
  \includegraphics[width=1.0\columnwidth]{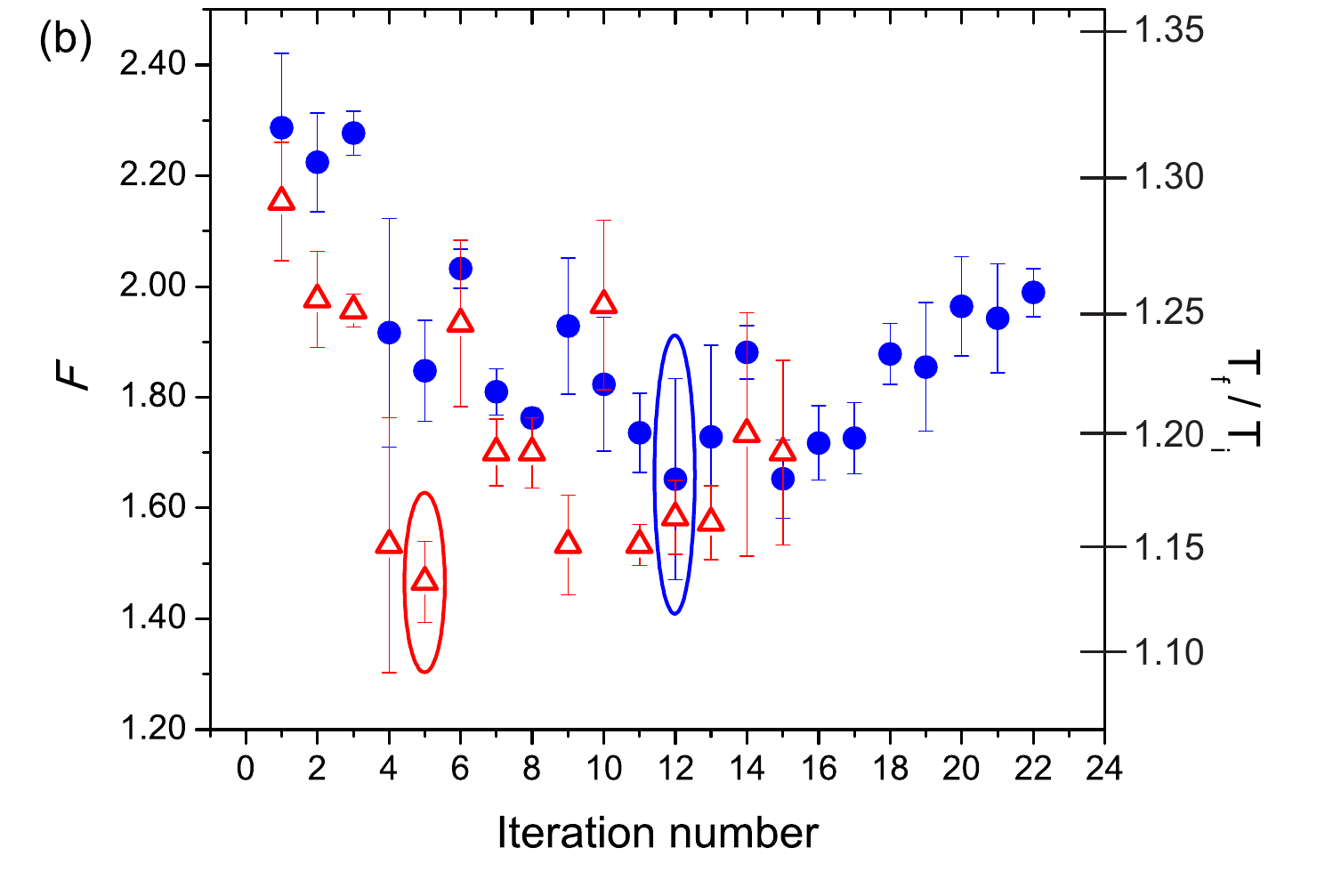}
  \label{fig:2DolRUN12fom}
}
\subfigure
{
  \begin{tabular*}{0.33\textwidth}{|c|c|c|c|}
  \hline
    & $\Delta t$ (ms) & $\tau$ (ms) & $F$  \\
    \hline
	 quasi-ad. & 140 & 30 & $1.66\pm 0.02$ \\
	 $s_{0}$ & 15 & 3 & $2.30\pm 0.03$ \\
	 $s_{opt}$ RUN 1 & 154 & 35 & $1.73\pm 0.02$ \\
         $s_{opt}$ RUN 2 & 45 & 9 & $1.40\pm 0.06$ \\
    \hline
  \end{tabular*}
 \label{tbl:2DOptCtrl}
}
\caption{3D-1D crossover. \subref{fig:2DolRUN12params}: Two-dimensional mapping (blu-colored palette) of the FOM as a function of the parameters $\Delta t$ and $\tau$. Circles (triangles) show the progress during the first (second) run of optimization (grey-colored palette). \subref{fig:2DolRUN12fom}: Measured figure of merit $F$ (ratio between final and initial thermal fraction of the atomic sample) after ramping up and down a two-dimensional optical lattice during the optimization loop. First (second) experimental run (see text) is represented by blue circles (red triangles). 
The two best results $F_{opt}$ measured in the two runs are circled in evidence. For completeness, on the right side of the graph we report also the corresponding values of the ratio between temperature $T_{f}$ measured after having switched on and off the lattices and temperature $T_{i}$ measured before loading the lattices (inferred from the thermal fraction assuming thermal equilibrium).
 The table reports the values of the FOM corresponding respectively to the quasi-adiabatic loading, to the exponential initial loading ($s_{0}$) and to the best one ($s_{opt}$) of the two runs, together with the correspondent values of the parameters $\Delta t$ and $\tau$. }
\label{fig:2DolRUN12}
\end{figure}

\textit{3D-1D crossover} -- We first consider the transformation between a three-dimensional BEC and an array of one-dimensional quasi-condensates obtained loading the 
BEC in a two-dimensional lattice. As a warm-up for the full CRAB optimization, here we optimize the process over a restricted class of functions, namely 
loading ramps of exponential shape with different duration $\Delta t$ and time constant $\tau$ as defined in Eq.\eqref{eq:ExpRamps}. The final value of the two lattices intensity $s_{max}=32$ is high enough to produce an array of independent one-dimensional gases where the transverse degrees of freedom are completely frozen and the tunneling rate of atoms between different sites is negligible on the time scale of the experiment. The thermalization time is $\Delta t_{th}=1\,\text{s}$. Before running the optimization algorithm, as the final thermal fraction is a two-dimensional function of the free parameters ($\Delta t, \tau$) and thus easily representable, we plotted it in the two-dimensional parameter phase-space (colored map in Fig.~\ref{fig:2DolRUN12params}). 
This allows the results of the closed-loop optimization to be compared with a brute-force approach, that is an extensive search in the parameter space. The colored map gives us interesting information on the problem structure, notice for example that for very short ramps the FOM is clearly higher than in case of longer ones, as expected. The extensive mapping approach is unfeasible as soon as the number of the parameters increases, however in this case where only two parameters are considered we show that the optimization finds very quickly the minima. Fig.~\ref{fig:2DolRUN12} presents two tests of the optimization loop characterized by the same initial guesses $\Delta t_{0}=15\,\text{ms}$ and $\tau_{0}=3\,\text{ms}$: $F$ is reported for both runs in Fig.~\ref{fig:2DolRUN12fom} as a function of the iteration number $n$ of the optimization loop. The possibility of finding different final results is due to the fact that performing an experiment characterized by a finite number of iterations and experimental errors, we may bump into little deviations in the measurements from run to run. 
As it can be seen, in both cases after a quick convergence to a minima the algorithm looks for other solutions possibly present in case the first were a local minima, but it founds none. 
In Fig.~\ref{fig:2DolRUN12params} the trajectories in the space of parameters clearly reflect this behavior. Finally, as reported in the table in Fig.~\ref{fig:2DolRUN12}, the two loops give two pairs of 
different final values $\left(\Delta t_{opt},\,\tau_{opt}\right)$: the second run ended in an improved results while the first missed the global minima, spending the available iterations in 
the large region where the FOM is almost flat, as we can see in Fig.~\ref{fig:2DolRUN12params}. 

\begin{figure}[t!]
\centering
\includegraphics[width=1\columnwidth]{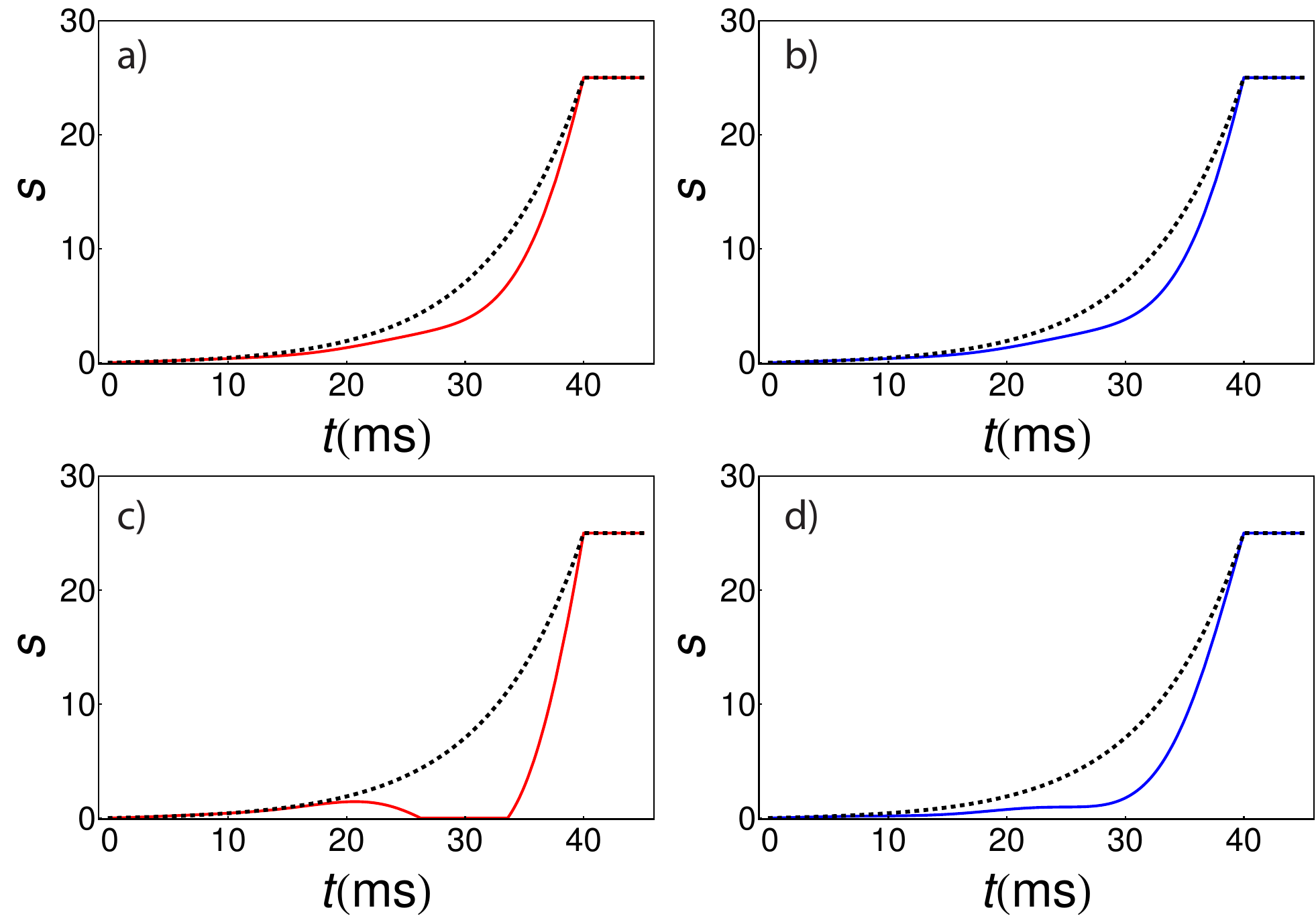}
\caption{Optical lattice ramps for the Superfluid - Mott insulator transition. Solid lines: a) First run initial guess. b) First run optimized ramp. c) Second run initial guess. d) Second run optimized ramp. The dashed black line represents the exponential non-corrected ramp.}
\label{fig:ramps11aprRUN12}
\end{figure}



\begin{figure}[t!]
\centering
\subfigure
{
\includegraphics[width=1\columnwidth]{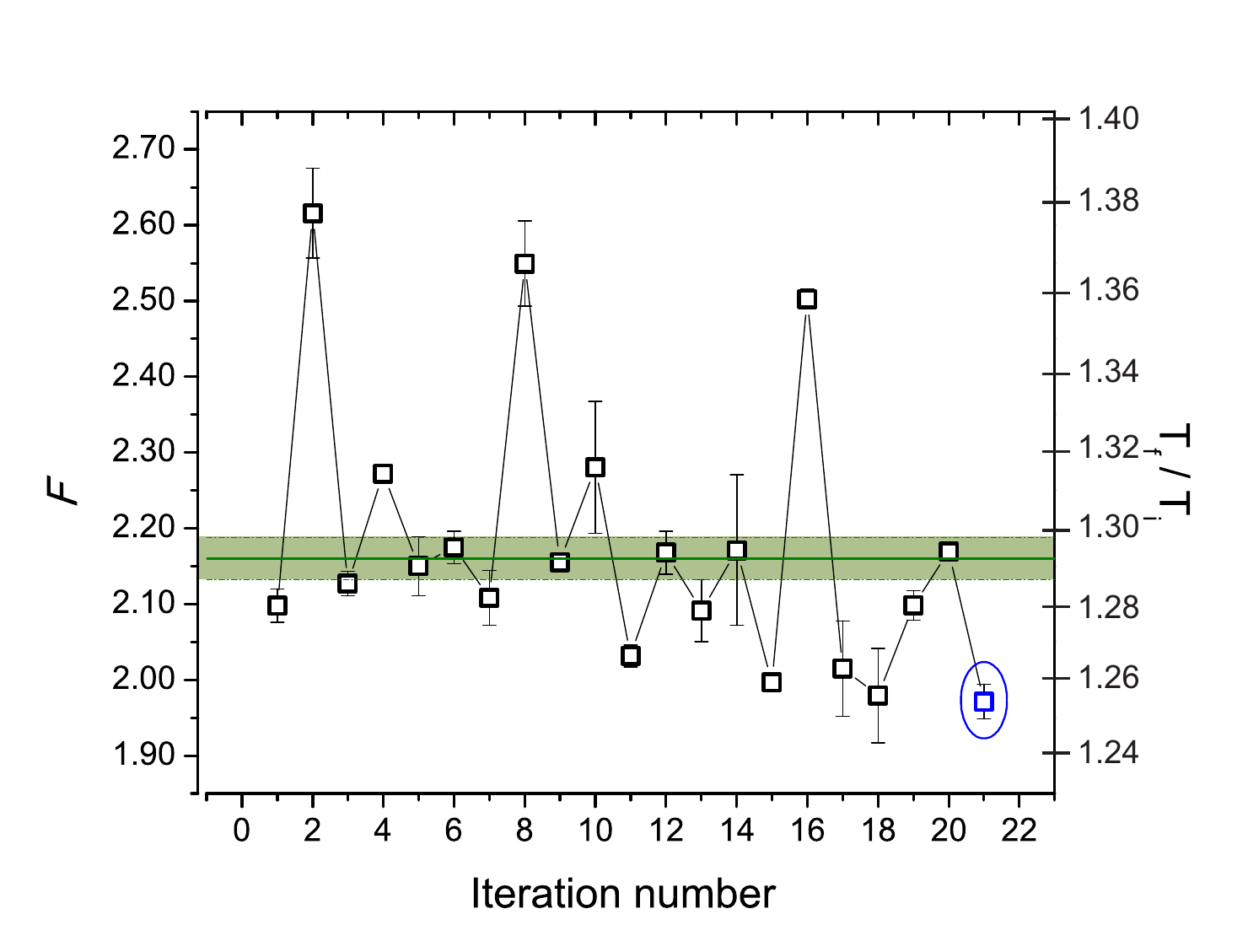}
}
\subfigure
{
  \begin{tabular*}{0.46\textwidth}{|c|c|c|c|c|c|}
    \hline
    & $\tau$ & $\Delta t$ & $[a_{1},b_{1},a_{2},b_{2}]$ & $F$ \\
    & (ms) & (ms) &  &  \\
    \hline    
    quasi-ad. & 140 & 30 & [0,0,0,0] & $2.16\pm 0.03$ \\
    $s_{uc}$ & 40 & 8 & [0,0,0,0] & $2.19\pm 0.03$ \\
    $s_{opt}$ RUN 1 & 40 & 8 & [0.2,0.2,0.1,0.1] & $1.89\pm 0.03$ \\
    $s_{opt}$ RUN 2 & 40 & 8 & [-0.09,-0.22,0.70,0.13] & $1.97\pm 0.02$ \\ 
    \hline
  \end{tabular*}
}

\caption{Optimization of the Superfluid - Mott insulator transition. 
The FOM $F$ is reported as a function of the iteration number $n$ (second run). The green region shows the FOM $F_{ad}=(2.16\pm 0.03)$ in the case of the quasi-adiabatic loading $\left(\Delta t=140\,\text{ms},\tau=30\,\text{ms}\right)$.
The blue-circled empty square reports the final optimal loading $F_{opt}$. In the table are reported the values of the parameters $[a_{1},b_{1},a_{2},b_{2}]$ of the correction and the FOM $F$ corresponding respectively to the quasi-adiabatic loading, to the exponential uncorrected loading ($s_{uc}$) and to the optimal loading ($s_{opt}$) found in the two runs.}
\label{fig:fom11aprRUN2}
\end{figure}

\textit{Superfluid-Mott Insulator transition} -- We now apply the closed-loop optimization to a more complex dynamical process, that is, the loading of the BEC into a three-dimensional lattice, effectively optimally driving the quantum phase transition from a superfluid to a Mott insulator phase. We perform a full CRAB optimization, that is, we search for the best possible correction of the form introduced in Eq.~\eqref{CRAB} to an initial exponential ramp $s_0(t)$ given in Eq.~\eqref{eq:ExpRamps}. Here we use two frequency components, i.e., we perform the optimization in a four-dimensional parameter space, with $\nu_{1}=1/\Delta t$ e $\nu_{2}=2/\Delta t$, where $\Delta t=40\,\text{ms}$ and $\tau=8\,\text{ms}$. The final optical lattice depth is $s_{max}=25$ -- deeply in the Mott insulating phase -- and the thermalization time $\Delta t_{th}=200\,\text{ms}$. As in the previous case, the FOM is represented by the normalized thermal fraction $F$, observed after an adiabatic switching off of the lattices, which measures the residual excitations on top of the Mott ground state. In this case, the quasi-adiabatic ramp $\left(\Delta t=140\,\text{ms},\tau=30\,\text{ms}\right)$ gives $F_{ad} \sim 2.16$ (green line in Fig.~\ref{fig:fom11aprRUN2}).
We have performed two different optimization runs characterized by different sets of initial random parameters $\{h_j\}$: In the first one the initial ramp 
is very close to the guess $s_0(t)$, while in the latter it is characterized by a very strong correction. The resulting optimal ramps are shown in Fig.~\ref{fig:ramps11aprRUN12}, together with the initial ramp and the uncorrected (pure exponential) one. It may be interesting to note that, despite starting from very different initial guesses, the two loops result 
in two very similar output ramps, characterized by a slow variation in the first part and a fast increase in the second part. 
The trend of $F$ during the second run is shown in Fig~\ref{fig:fom11aprRUN2}. The results of the two optimizations 
are summarized in the table. 
Note that, although the very small number of parameters involved in the optimization ($n_f=2$), we obtain a transformation about three times faster than the adiabatic one and with a final temperature improved by a few percent. Whenever additional improvements would be needed for real applications, where improved precisions and/or faster ramps are crucial elements for a successful experiment -- for example to reduce decoherence effects -- the presented optimization can be improved adding more optimization parameters and/or 
increasing the optimization runs.

\textit{Conclusions} --  We have shown for the first time that closed-loop optimal control can be effectively applied to the manipulation of cold atoms in optical lattices, demonstrating how it is possible to develop strategies different from adiabatic ones. In particular we have shown that it is possible to drive the 3D-1D crossover and the 3D Superfluid-Mott insulator transition over time scales about one third shorter than a standard quasi-adiabatic one, while still improving on the final state. Our strategy can be applied to the implementation of the ground state of different Hamiltonians. This demonstration paves the way to further 
developments that will allow to face the challenges of the next generation experiments with cold atoms in optical lattice.

\textit{Acknowledgements} --  We acknowledge support from the EU IP-project AQUTE and the project EU-STREP MALICIA. 
The work at LENS has been supported by MIUR through PRIN2009, ERC Advanced Grant DISQUA, IIT Seed Project ENCORE.
TC and SM acknowledge support from the DFG via the SFB/TRR21.


\begin{thebibliography}{99}
\bibitem{noteFreq} The characteristic frequencies of the harmonic trapping experienced by atoms are about ($40\,\text{Hz}\times 9\,\text{Hz}\times 40\,\text{Hz}$).
\bibitem{noteThFr} Note that the measurement is performed after the lattice switching-off, since we do not have access to direct information on temperature and/or excitations.
\end{thebibliography}
\end{document}